\documentclass[%
 reprint,
superscriptaddress,
 amsmath,amssymb,
 aps,
]{revtex4-2}
\usepackage{graphicx}
\usepackage{dcolumn}
\usepackage{bm}
\usepackage{comment}
\usepackage{braket}

\usepackage{hyperref}
\usepackage{xcolor}
\definecolor{cite}{rgb}{0.,0.,0.9}  
\hypersetup{colorlinks,linkcolor={cite},citecolor={cite},urlcolor={cite}}

\begin{document}


\title{Experimental and theoretical study of dynamic polarizabilities in the \texorpdfstring{$5S_{1/2}\,-\,5D_{5/2}$}{5S1/2-5D5/2} clock transition in rubidium-87 and determination of E1 matrix elements}

\author{Rhona Hamilton}
\affiliation{
 Institute for Photonics and Advanced Sensing (IPAS) and School of Physical Sciences, University of Adelaide, Adelaide SA 5005, Australia
}
\author{Benjamin M. Roberts}
\affiliation{
School of Mathematics and Physics, The University of Queensland, Brisbane QLD 4072, Australia
}
\author{Sarah K. Scholten}
\email{sarah.scholten@adelaide.edu.au}
\affiliation{
 Institute for Photonics and Advanced Sensing (IPAS) and School of Physical Sciences, University of Adelaide, Adelaide SA 5005, Australia
}

\author{Clayton Locke}
\affiliation{
 Institute for Photonics and Advanced Sensing (IPAS) and School of Physical Sciences, University of Adelaide, Adelaide SA 5005, Australia
}
\author{Andre N. Luiten}
\affiliation{
 Institute for Photonics and Advanced Sensing (IPAS) and School of Physical Sciences, University of Adelaide, Adelaide SA 5005, Australia
}
\author{Jacinda S. M. Ginges}
\affiliation{
School of Mathematics and Physics, The University of Queensland, Brisbane QLD 4072, Australia
}
\author{Christopher Perrella}
\affiliation{
 Institute for Photonics and Advanced Sensing (IPAS) and School of Physical Sciences, University of Adelaide, Adelaide SA 5005, Australia
}

\date{\today}

\begin{abstract}
The interaction between light and an atom causes perturbations in the atom's energy levels, known as the light-shift. These light-shifts are a key source of inaccuracy in atomic clocks, and can also deteriorate their precision. 
We present a study of light-shifts and associated dynamic polarizabilities for a two-photon atomic clock based on the $5S_{1/2}$\,--\,$5D_{5/2}$ transition in rubidium-87 over the range 770\,nm to 800\,nm. We determine experimental and theoretical values for a magic wavelength in this range and the electric dipole (E1) matrix element for the $5P_{3/2}$\,--\,$5D_{5/2}$ transition. 
We find a magic wavelength of 776.179(5)\,nm (experimental) and 776.21\,nm (theoretical) in the vicinity of the $5P_{3/2}$\,--\,$5D_{5/2}$ resonance, and the corresponding reduced E1 matrix element $1.80(6)\,ea_0$ (experimental) and $1.96(15)\,ea_0$ (theoretical). These values resolve a previous discrepancy between theory and experiment.

\end{abstract}

\maketitle


\section{Introduction} \label{sec:level1}
High-precision frequency references underpin a range of military, government, and private position, navigation, and timing (PNT) applications in modern society such as the global positioning system (GPS)\,\cite{Knappe_2004, ESNAULT2011854, Batori2021}. 
Further use of such frequency references is also found in fundamental tests of physics using atoms and molecules, tests of relativity, and astronomy\,\cite{Safronova2018, Grotti2018, Doeleman2011}.

For many of these applications, the limits on factors such as range, sensitivity and accuracy can be improved by using a more precise and accurate timing signal\,\cite{Safronova2018, Grotti2018, Doeleman2011, Batori2021}.
However, these applications also require that the compact clock supplying the timing signal must maintain this precision when deployed outside of the laboratory. These requirements motivate a clock design that is robust to disturbances and environmental variations while minimizing size, weight and power (SWaP). 

The $5S_{1/2}$\,--\,$5D_{5/2}$ transition in rubidium has emerged as a promising candidate for an atomic frequency standard that meets these requirements\,\cite{Poulin1997,Perrella2013,Martin2018,Gerginov2018, Perrella2019, Newman2019, Maurice2020, Newman2021, Lemke2022}. 
This optical transition lends itself well to this application because it has a number of convenient properties. 
The two-photon excitation scheme allows counter-propagating optical beams to be used, producing a very narrow transition that may be free from Doppler broadening\,\cite{Grynberg_1977}. 
The excitation of this two-photon process is particularly efficient because of the intermediate $5P_{3/2}$ state, which falls almost half way between the $5S_{1/2}$\,--\,$5D_{5/2}$ transition of interest, leading to high signal-to-noise ratios. 

The two-photon transition can be excited by either two 778\,nm photons \cite{Martin2018} or a 780\,nm and 776\,nm photon \cite{Perrella2013}, which is the scheme used in this work. The advantage of the single-color excitation scheme is that the transition is completely free from Doppler broadening \cite{Martin2018}. The two-color excitation has the advantage of being much more efficient in driving the two-photon transition, at the expense of residual Doppler broadening due to the mismatch in photon energies \cite{Perrella2013}. 

A major factor limiting the accuracy and precision of atomic clocks, such as this rubidium two-photon clock, is the presence of light-shifts\,\cite{Takamoto2005, Martin2019, Gerginov2018}.
Light-shifts arise from the interaction between the electromagnetic field of the laser(s) used to excite the clock transition with the atom (see for e.g.\,\cite{Mitroy2010}).
This interaction perturbs the energy levels of the clock transition, which in turn shifts the clock frequency.
This shift is dependent on both the frequency and intensity of the perturbing light fields.
Therefore, fluctuations in the laser powers result in a reduction in the stability of the clock\,\cite{Arora2007b}. 
On the other hand, to maximize the signal-to-noise of the clock readout it is desirable to use the highest possible optical powers to excite the clock transition, particularly in the case of a two-photon transition. 
Therefore, there is a trade-off between maximizing the signal-to-noise ratio and reducing deleterious light-shifts.

Recent work has shown it is possible to mitigate these light-shifts in a two-photon rubidium atomic clock\,\cite{Gerginov2018, Martin2019, Lemke2022}. 
For the single-color excitation at 778\,nm, a secondary off-resonance laser could be applied to reduce the effect of light-shifts on the clock transition\,\cite{Martin2019}. 
While for the two-color two-photon excitation process, balancing of the excitation laser powers has been shown to mitigate light-shifts on the clock transition\,\cite{Gerginov2018}. 
The two-color two-photon excitation process allows for the possibility of tuning the wavelengths, intensities, and polarizations of the two beams such that the light-shift induced by one beam exactly cancels that of the other. 
In order to take advantage of this principle in future clock designs, a detailed understanding of the underlying atomic physics is required. 
In particular, we seek accurate values of the electric dipole (E1) matrix elements and dynamic polarizabilities, which determine the strength of the light-shifts associated with the relevant clock transitions.

In this paper we present both the experimentally- and theoretically-determined differential dynamic polarizabilities which show excellent agreement across the explored wavelength range from 770\,nm to 800\,nm. 
We find a magic wavelength of 776.179(5)\,nm (experimental) and 776.21\,nm (theoretical) in the vicinity of the $5P_{3/2}$\,--\,$5D_{5/2}$ resonance, and the corresponding reduced E1 matrix element $1.80(6)\,ea_0$ (experimental) and $1.96(15)\,ea_0$ (theoretical), where $e$ the elementary charge, and $a_0$ is the Bohr radius.
This resolves a previous discrepancy between theoretical~\cite{Safronova2011} and experimental~\cite{Whiting2016,*Whiting2018} results for this transition.

This paper is structured in the following way. In Section~\ref{sec:difPolBG} we present the theory background for the differential polarizability. In Section~\ref{sec:experiment} we describe the experimental setup for the measurements, and in Section~\ref{sec:LS2difPol} we show how the conversion from measured light-shifts to differential dynamic polarizabilities of the clock transition is made. The method for the atomic many-body theory calculations is described in Section~\ref{sec:theory}. Finally, the experimental and theoretical results are presented and discussed in Section~\ref{sec:Res}, and our conclusion is presented in Section~\ref{sec:conclusion}.

\section{Differential Polarizability Background} \label{sec:difPolBG}
The perturbation of atomic energy levels due to the presence of an oscillating electric field is known as a light (or AC Stark) shift. For an optical clock, the clock frequency will be affected by light-shifts to the ground and excited states. As such, the light-shifts discussed in this work refer to the difference between the shifts experienced by the two energy levels in the clock transition.
In the following description connecting the light-shift to dynamic polarizabilities we follow Ref.~\cite{Arora2007b}; atomic units are used in this section ($e\,{=}\,\hbar\,{=}\,m_e\,{=}\,1$). 

The light-shifts induced by an oscillating external electric field $\boldsymbol{\epsilon}$ arise due to mixing of opposite-parity states by the interaction operator $V_I$, given by 
\begin{equation}
    V_I = -\boldsymbol{\epsilon}\cdot\mathbf{d}\, ,
\end{equation}
where \textbf{d} is the electric dipole operator. 
The first-order correction to the energies vanishes, and the lowest-order correction to the energy of the atom in state $v$ is given by
\begin{equation}
    \delta E = \sum_k \frac{\langle v|V_I|k\rangle \langle k|V_I|v\rangle}{E_v - E_k}\,  ,
\end{equation}
where $v$ and $k$ are unperturbed atomic states with energies $E_v$ and $E_k$ respectively. 
The sum is taken over all intermediate states $k$ permitted by the selection rules.
The energy shift of state $v$, due to a time-varying electric field with frequency $\omega$, is proportional to the dynamic polarizability of the state $\alpha(\omega)$,
\begin{equation}
    \delta E = -\frac{1}{2}|\boldsymbol{\epsilon}|^2\alpha(\omega)\, .
\end{equation}
The light-shift of a transition, $\Delta E$, is the difference between the perturbation in energy experienced by the two levels in the transition,
\begin{equation} \label{eqn:EnergyLightShift}
    \Delta E = -\frac{1}{2}|\boldsymbol{\epsilon}|^2 \Delta\alpha(\omega)\, ,
\end{equation}
where $\Delta\alpha(\omega)$ is the differential polarizability between the ground, $g$, and excited, $e$, states of the affected transition, 
\begin{equation}\label{eq:DeltaAlpha}
    \Delta\alpha(\omega) = \alpha_{g}(\omega) - \alpha_{e}(\omega)\, .
\end{equation}

The dynamic polarizability of the state $v$ may be expressed in terms of scalar and tensor components, $\alpha_0(\omega)$ and $\alpha_2(\omega)$, 
\begin{equation} \label{eq:pol}
   \alpha (\omega) = \alpha_0(\omega) + \alpha_2(\omega)\frac{3m_v^2-J_v(J_v+1)}{J_v(2J_v - 1)} \, ,
\end{equation}
where $J_v$ and $m_v$ are the total angular momentum and its projection (see, e.g., Ref.~\cite{Arora2007b,Rosenbusch2009,kien2013}).  
The scalar polarizability is given by 
\begin{equation} \label{eq:matElemScal}
    \alpha_0(\omega) = \frac{2}{3(2J_v+1)}\sum_{k}\frac{\omega_{kv}|\langle v||d||k\rangle|^2}{\omega^{2}_{kv}- \omega^2}\, ,
\end{equation}
where $\langle v||d||k\rangle$ are reduced E1 matrix elements and $\omega_{kv}$ are  transition frequencies. Similarly, the tensor polarizability is expressed in terms of the reduced E1 matrix elements in the following way: 
\begin{equation} \label{eq:matElemTens}
    \alpha_2(\omega) = C\sum_{n}(-1)^{J_v+J_k} \begin{Bmatrix} J_v & 1 & J_k \\ 1 & J_k & 2\end{Bmatrix}
    \frac{\omega_{kv}|\langle v||d||k\rangle|^2}{\omega^{2}_{kv}-\omega^2}\,,\\
\end{equation}
where
\begin{equation} 
    C=4\left(\frac{5J_v(2J_v-1)}{6(J_v+1)(2J_v+1)(2J_v+3)}\right)^{1/2} \, ,\nonumber
\end{equation}
and $\{:::\}$ is a 6$j$-symbol. 
The polarizabilities are calculated using all-orders relativistic many-body methods described in Section~\ref{sec:theory}.

\section{Experimental Setup} 
\label{sec:experiment}

\begin{figure}[t]
    \centering
    \includegraphics[width=\columnwidth]{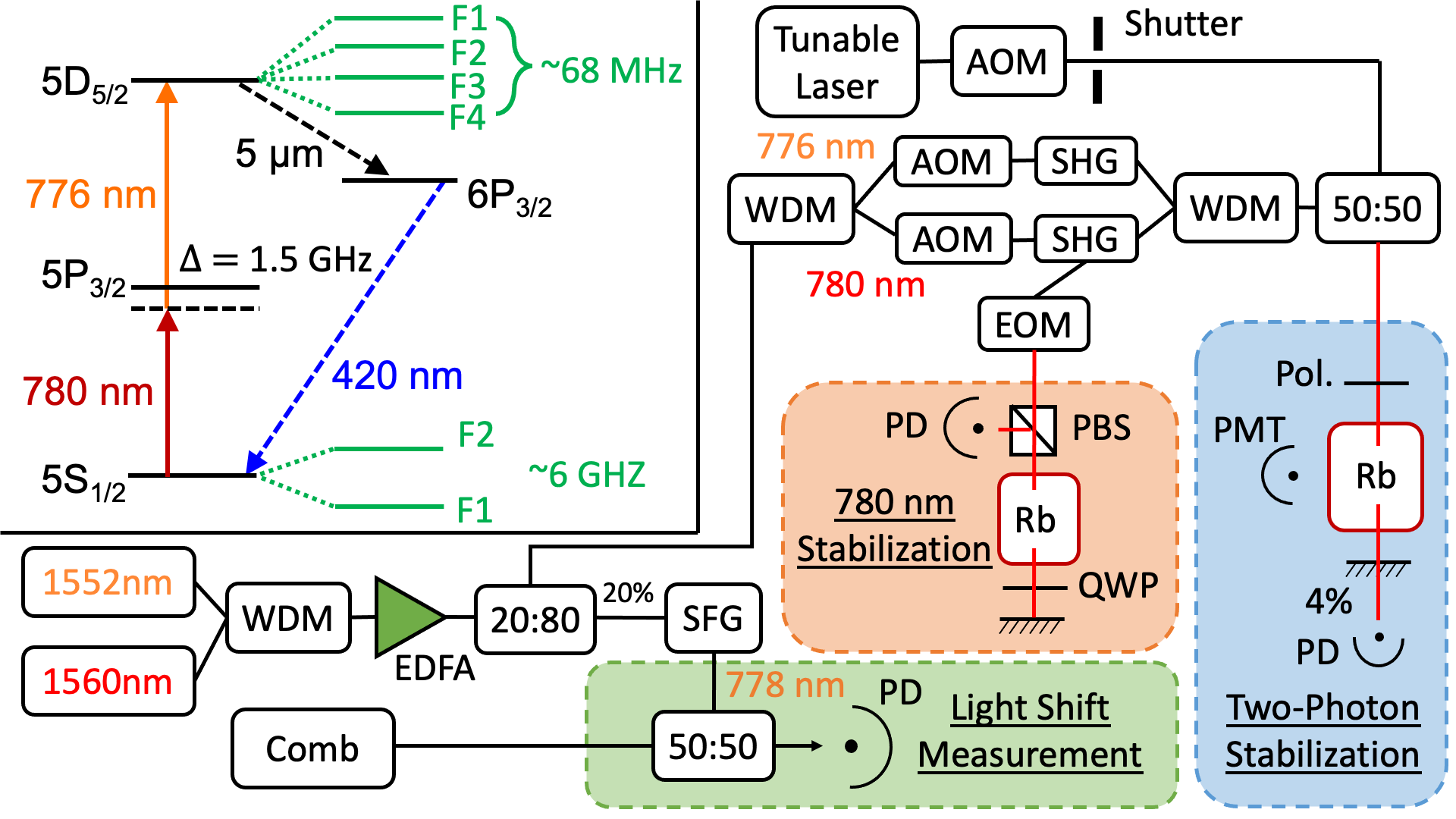} 
    \caption{Relevant energy levels in rubidium-87, and experimental schematic of the clock modified to measure light-shift. AOM: acousto-optic modulator, SHG: second harmonic generator, WDM: wavelength division multiplexer, EDFA: erbium doped fiber amplifier, SFG: sum frequency generator, EOM: electro-optic modulator, PD: photo-diode, PBS: polarization beam splitter, QWP: quarter wave plate, PMT: photo multiplier tube, BGF: blue glass filter, Pol: linear polarizer.}
    \label{fig:levels_setUp}
\end{figure}

 The experimental schematic of the clock, and the adjustments made to measure light-shifts, are presented in Fig.\,\ref{fig:levels_setUp}, with a list of experimental parameters summarized in Table\,\ref{tab:exp_param}. 
 The fundamental operation of the clock is described in Ref.~\cite{Perrella2019}, and we present a short description here. 
 The lasers driving the clock are 1552\,nm and 1560\,nm fiber lasers.
 These lasers are combined and amplified through an Erbium Doped Fiber Amplifier (EDFA), and frequency doubled to 776\,nm and 780\,nm to excite the two-photon transition between the $5S_{1/2}(F\,{=}\,2)$ and $5D_{5/2}(F\,{=}\,4)$ levels in rubidium-87 \cite{Nez1993}, shown in Fig.\,\ref{fig:levels_setUp}. 
 The 780\,nm laser is frequency shifted by $\Delta\,{=}\,1.5$\,GHz using an electro-optic modulator (EOM) and stabilized to a saturated absorption spectroscopy feature of the $5S_{1/2}(F\,{=}\,2)\,{-}\,5P_{3/2}(F\,{=}\,1)$ transition. 

The 776\,nm laser is stabilized to the $5S_{1/2}(F\,{=}\,2)\,{-}\,5D_{5/2}(F\,{=}\,4)$ two-photon transition of rubidium-87 via the 420\,nm fluorescence produced by the $6P_{3/2}\,{-}\,5S_{1/2}$ transition which is detected by a photo-multiplier tube (PMT). 
The 776\,nm frequency lock stabilizes the laser to maximize the 420\,nm fluorescence extracted from the rubidium atoms. 
The two-photon transition is excited within a rubidium cell heated to 80$^\circ$C to increase the rubidium vapor density, thus the fluorescence power detected and improve the short-term stability of the clock. 
All optical fields are polarized by a linear polarizer before passing through the rubidium cell.
The transition is excited by a retro-reflected beam of two co-propagating laser beams at slightly different wavelengths, and this difference results in a decrease in the effectiveness of Doppler compensation, yielding a transition width of $3.7\,{\pm}\,0.1$\,MHz ($1/e\,{\approx}\,0.368$ full width).
Frequency modulation spectroscopy is employed, implemented via an acousto-optic modulator (AOM) acting upon the 776\,nm laser, to produce a feedback signal to stabilize the sum of the 780\,nm and 776\,nm lasers to the two-photon transition.
The AOM's also provide a means to power stabilize the 780\,nm and 776\,nm lasers at the Rb cell in which the two-photon transition is excited.
 
The stabilized clock output is the sum-frequency of the 1552\,nm and 1560\,nm lasers at 778\,nm which is compared to a commercial frequency comb stabilized to an ultra-low expansion (ULE) cavity as a frequency reference. The stability of the clock over the time frame of a measurement ($\approx$ 250\,s) is approximately 300\,Hz$_{\textrm{rms}}$.

To produce the light-shifts to be measured, a third tunable laser is combined with the 780\,nm and 776\,nm clock lasers via a polarization maintaining (PM) 50:50 fiber coupler prior to launching into the heated rubidium cell.
Light-shifts induced by the tunable laser perturb the energy levels associated with the two-photon transition. 
Thus, the frequency associated with the maximum fluorescence shifts, and the clock output frequency tracks this change. 
These frequency changes, and hence the light-shifts, are measured through comparison of stabilized 778\,nm output to the frequency comb. 

A titanium-sapphire (Ti:Sapph) laser that could be tuned between 760\,nm and 800\,nm was used to induce light-shifts. 
The wavelength of the tunable laser for each light-shift measurement was measured using a wavemeter (accuracy $\pm$200\,MHz, precision $\pm$5\,MHz).

Light-shift measurements were made in a differential fashion, whereby the Ti:Sapph laser periodically illuminated the rubidium cell, enabling a difference measurement to be made with and without the induced light-shifts. 
This enabled rejection of any drifts of the clock frequency over time. 
The measurement consisted of a repeated sequences of $\approx 25$\,s of exposure from the Ti:Sapph laser followed by the same period without exposure, which was implemented using an optical shutter. 
While on, the Ti:Sapph laser induced light-shifts dependent on the optical power of the laser beam, which was controlled by an AOM.

For each frequency step of the tunable laser, the induced variation in the clock frequency was measured as a function of the tunable laser's power. 
The resulting light-shift per Watt (Hz/W) is extracted for each wavelength of the  tunable laser which is converted to differential polarizability, see Sec.\,\ref{sec:LS2difPol}. 

The largest source of uncertainty in the light-shift measurements is our measurement of the optical power of the Ti:Sapph laser interacting with the rubidium atoms. The optical power was measured using a silicon photodiode (New Focus 2031, low gain) located behind the Rb cell and retro-reflection mirror (4\% transmission). 
The output of the New Focus photodiode was calibrated using a calibrated power meter (Thorlabs S120C Silicon photodiode). This calibration was repeated 15 times at differing optical powers in order to estimate the uncertainty of the power measurement. This gave a photodiode sensitivity of 0.55 $\pm$ 0.05 A/W, from which the relative uncertainty in the power measurement is $\pm 10\%$, which dominates our experimental error. 

\begin{table}[htb]
    \caption{Nominal values for experimental parameters and associated absolute uncertainties. The beam diameter is quoted as the $1/e^2\,{\approx}\,0.135$ relative intensity diameter. Absolute uncertainties in optical powers are estimated from the 10\% uncertainty in the photodiode sensitivity. Natural line width (full width at half maximum), and associated uncertainty, of the $5D_{5/2}$ state is derived from the lifetimes in Ref.~\cite{Sheng2008}.}
    \label{tab:exp_param}
    \begin{ruledtabular}
    \begin{tabular}{lcc}
    Parameter & Nom. Val. & Abs. Unc. \\
    \hline
     780\,nm laser power (mW)  & 0.75 & 0.08\\
     776\,nm laser power (mW) & 1.2 & 0.1 \\
     Beam diameter (mm) & 1.5 & 0.1 \\
     Detuning from $5P_{3/2}(F=3)$ (GHz)& 1.50 & $\pm0.01$\\
     Cell temperature ($^\circ$C) & 80 & $\pm10$ \\
     Residual Doppler broadening (MHz)& 3.7 & $\pm$0.1 \\
     $5D_{5/2}$ natural line width (kHz) & 667 & 6
    \end{tabular}
    \end{ruledtabular}
\end{table}

\section{Experimental Determination of Differential Polarizability}
\label{sec:LS2difPol}

To extract the differential polarizability from light-shift measurements, we need to understand how the light-shift measurement averages over the Gaussian beam intensity profile of the perturbing laser.
To achieve this we model the light-shifted two-photon fluorescence spectra by averaging over the driving and perturbing laser beam profiles. 

We model the two-photon transition's spectral profile as a Gaussian lineshape, as the underlying natural Lorentzian lineshape is masked by residual Doppler broadening.
We use the measured $1/e$ full-width of the fluorescence spectra, $\sigma\,{=}\,3.7(0.1)$\,MHz, in the model; see Table~\ref{tab:exp_param}.
The two-photon fluorescence spectral lineshape at a given radial point of the beam profiles, $F(\Delta_{776}, r)$, has an amplitude proportional to the product of the 780\,nm and 776\,nm exciting laser intensities, $I(r)_{780}$ and $I(r)_{776}$, respectively\,\cite{Bjorkholm1976, Perrella2018}. 
The resonant frequency of the two-photon transition experiences a light-shift induced by the perturbing Ti:Sapph laser.
To account for the spatial dependence of the laser beam, we integrate over the beam area to produce an average fluorescence spectral line shape, $F(\Delta_{776})$, for a given 776\,nm laser detuning, $\Delta_{776}$:
\begin{equation}
    F(\Delta_{776}) =  \int_0^\infty F(\Delta_{776}, r)\,r\,dr,
\end{equation}
with,
\begin{equation}
    F(\Delta_{776}, r) =  I(r)_{780}I(r)_{776}\,e^{\displaystyle \small-\left(\frac{\Delta_{776} - \Delta\nu(r,\omega)}{\sigma}\right)^2} , \notag
\end{equation}
where the Gaussian term represents the spectral lineshape of the light-shifted two-photon transition, and the $r$ factor originates from the Jacobian in cylindrical coordinates.
We assume the intensity profiles do not change over the length of the rubidium vapor cell as the two-photon transition absorbs minimal light from any of the laser beams. 

The light-shift of the two-photon transition at a given radial point in the beam is given by:   
\begin{equation} \label{eqn:LSrdep}
    \Delta\nu(r, \omega) = \frac{\Delta\alpha(\omega)}{2\,c\,\varepsilon_0\,h}I(r)_\textrm{Ti}\, ,
\end{equation}
where $I(r)_\textrm{Ti}$ is the radial intensity profile of the Ti:Sapp laser which induces the lights shifts, and $\omega$ is its angular frequency.
This expression is found from Eq.~\eqref{eqn:EnergyLightShift} by converting energy to linear frequency via $E\,{=}\,h\nu$, by expressing the laser's electric field in terms of intensity via $I\,{=}\,|\mathbf{S}|\,{=}\,\varepsilon_0 c |\boldsymbol{\epsilon}|^2$, and by converting from atomic units to SI units using the factor $4\pi\varepsilon_0\mathrm{a}_0^3$.
Here, $\varepsilon_0$ is the vacuum permittivity, $c$ is the speed of light, and $\mathrm{a}_0$ is the Bohr radius.

\begin{figure}
    \centering
    \includegraphics[width=\columnwidth]{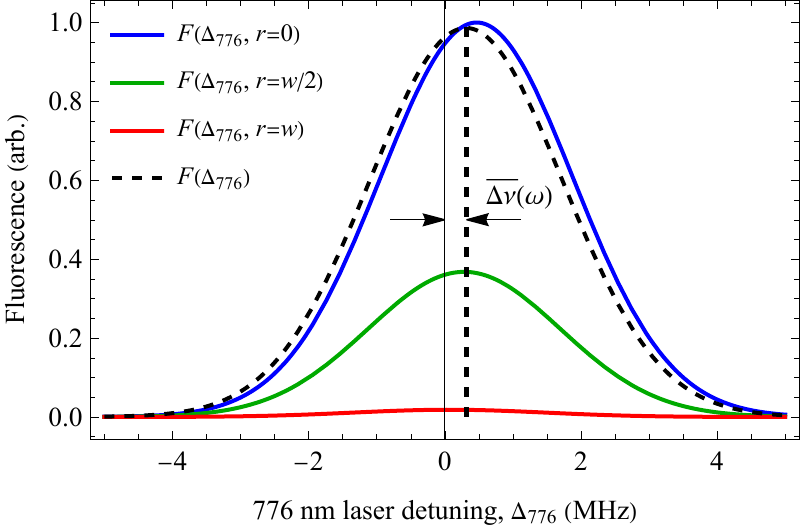}
    \caption{
    Modeled fluorescence spectrum as a function of 776\,nm laser detuning, $\Delta_{776}$.
    The dashed line is an example of an averaged spectra over the laser beam from which the value of light-shift $\overline{\Delta\nu}(\omega)$ is extracted. 
    Solid lines are fluorescence spectra at different radial positions within the laser beam of beam waist $w$ ($1/e^2$ intensity diameter).}
    \label{fig:distOfLS}
\end{figure}

We evaluate $F(\Delta_{776})$ over a range of 776\,nm laser detunings to produce a two-photon fluorescence spectrum, shown in Fig.\,\ref{fig:distOfLS}.
The frequency detuning that gives the maximum fluorescence is the averaged light-shift, $\overline{\Delta\nu}(\omega)$, as measured by the clock.
By simulating two-photon spectra at a range of Ti:Sapph powers, $P_\textrm{Ti}$, we are able to derive a conversion factor between the measured two-photon light-shift averaged over the laser beam profile at a given Ti:Sapph power, and the atomic differential polarizability.
This conversion factor is found to be approximately 3.6, yielding:
\begin{equation}
    \Delta\alpha(\omega) \approx \frac{1}{3.6}\frac{\overline{\Delta\nu}(\omega)}{P_\textrm{Ti}},
\end{equation}
where $\overline{\Delta\nu}(\omega)/P_\textrm{Ti}$ is extracted from our measurements.

\section{All-orders many-body calculations}
\label{sec:theory}

Rubidium has a single valence electron above a closed krypton--like core.
The natural starting point for the calculation is the relativistic Hartree-Fock (RHF) model.
A non-local energy-dependent operator, $\Sigma$, is then added to the RHF equation for the valence electron to account for the core-valence correlations~\cite{DzubaPNC1984,*DzubaPNC1985},   
\begin{equation}\label{eq:H-Bru}
(h_{\rm HF} +  \Sigma)\psi^{\rm (Br)} = \varepsilon^{\rm (Br)}\psi^{\rm (Br)},
\end{equation}
where $h_{\rm HF}$ is the single-particle RHF Hamiltonian. 
This yields correlated (``Brueckner'') energies and orbitals. 
To calculate $\Sigma$, we use the Feynman diagram technique~\cite{DzubaCPM1988pla,*DzubaCPM1989plaEn},
in which the dominating correlations (screening of the Coulomb interaction by the core electrons, and the hole-particle interaction inside hole-particle loops) are included to all orders in the Coulomb interaction. 
To gauge the importance of inclusion of the higher-order correlations, we also calculate correlations to lowest (second) order in the residual Coulomb interaction through use of the second-order correlation potential $\Sigma^{(2)}$.

\begin{figure*}[t]
\includegraphics[width=\textwidth]{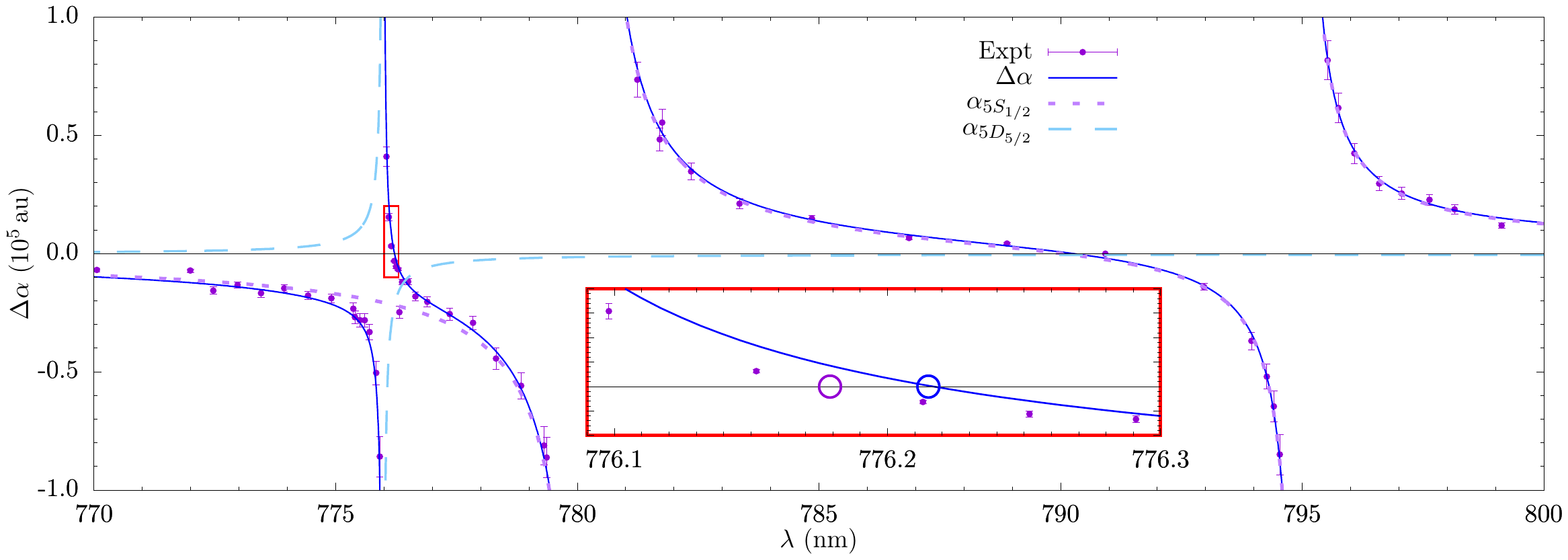}
\caption{\label{fig:results} 
Measured values for differential polarizability $\Delta\alpha(\lambda)$ [with $m=1/2$, see Eq.~\eqref{eq:pol}] for the $5S_{1/2}$\,--\,$5D_{5/2}$ transition of Rb (purple markers), overlaid with theoretical prediction (solid blue line).
The results of calculations for the individual $5S$ and $5D_{5/2}$ polarizabilities are also shown (dotted and dashed lines, respectively).
The inset shows the same, zoomed in to the region of the magic wavelength around 776\,nm; the purple and blue circles show the experimental and theoretical values for the magic wavelength, respectively.}
\end{figure*}

In the presence of an external laser field, the core electron wavefunctions are perturbed, 
$\psi \to \psi + \delta\psi$.
This leads to a correction to the RHF potential that contains the external field. This {\it core polarization} correction is an important addition to the E1 matrix elements. 
We account for this using the time-dependent Hartree-Fock method (TDHF)~\cite{DzubaHFS1984}, which is equivalent to the random phase approximation with exchange. 
By calculating the matrix elements using Brueckner orbitals and accounting for core polarization, the most important correlation effects are included.

There are also non-Brueckner correlation effects, 
the most important being the structure radiation, which arises from the perturbation of the correlation potential by the external field, and the normalization of states, which arises due to the change of the normalization of the many-body wavefunction~\cite{Dzuba1987jpbRPA}.
We account for these at the lowest (third) order of perturbation theory.
We also account for the Breit interaction (see, e.g.,~Ref.~\cite{BetheBook}) and quantum electrodynamics (QED) radiative corrections~\cite{Ginges2005}.
Finally, we include the effect of missed higher-order correlations through the introduction of semi-empirical scaling factors for the correlation potential, $\Sigma\to\lambda\Sigma$,
which are tuned to match the experimental energies.
The factors are very close to 1, due to the already high accuracy of the {\sl ab initio} results~\cite{Roberts2022}.
These semi-empirical corrections allow us to estimate the dominant contribution to the uncertainty in the theory calculations. This is discussed further in Ref.~\cite{Roberts2022}, where full details of the method of calculation may be found. 

To calculate the polarizabilities, a full spectrum of intermediate states is required, as may be seen from Eqs.~\eqref{eq:matElemScal} and \eqref{eq:matElemTens}.
To improve the numerical precision of the calculations, the sum over (discrete and continuous) intermediate states in the polarizability is reduced to a sum over a pseudospectrum of discrete states, by placing the atom in a cavity. 
The cavity is made large enough so that the pseudostates match the physical states for the lowest-lying levels.

As a test of the accuracy of the calculations, we calculate E1 matrix elements and scalar and tensor polarizabilities for the dynamic and static cases, and compare the results against available experimental and theoretical data.  
We present these results in the following section. 
Along with checks of the stability of the results, and estimates of the size of missed correlations and other effects, these comparisons provide an excellent gauge of the accuracy of the calculations. Our results demonstrate an accuracy on the level of 0.1\% for most of the E1 matrix elements.

\section{Results}\label{sec:Res}
In Section~\ref{sec:Res:MagicWav} we present the results of our measured and calculated differential polarizabilities, and we deduce -- both experimentally and theoretically -- the magic wavelengths for the $5S_{1/2}$\,--\,$5D_{5/2}$ transition for the first time. In Section~\ref{sec:Res:E1}, we present the results of E1 matrix elements extracted from our measured differential polarizabilities, alongside the results of our theoretically-determined matrix elements. 
This includes, most notably, the $5P_{3/2}$\,--\,$5D_{5/2}$ matrix element, for which our results agree, and they resolve a previous discrepancy between theory and experiment.  
These quantities are of particular interest for mitigating the light-shift in atomic clocks.

In Fig.~\ref{fig:results}, our experimentally- and theoretically-determined polarizabilities are presented, and the magic wavelengths are indicated. 
In Table~\ref{tab:alpha}, we present static ($\omega\,{=}\,0$) scalar and tensor polarizabilities calculated at various levels of many-body approximation, along with the results of other theory and experiments as a check of the accuracy of the methods. 
Our experimental and theoretical results for E1 matrix elements of relevance to the considered polarizabilities are shown in Table~\ref{tab:E1}, along with the results of other theory and experiments. 
The high accuracy of the calculations is seen from these data (see also Ref.~\cite{Roberts2022}).
Also seen from the tables, in the breakdown of the results, is the high sensitivity of the $5P$\,--\,$5D$ E1 matrix elements, and thus the polarizabilities that depend on them, to correlation corrections. We discuss the results further in the subsections below.

\subsection{Magic Wavelength Determination} \label{sec:Res:MagicWav}
The magic wavelength is defined as the wavelength at which the induced light-shifts in the two states of the considered transition are equal and cancel each other out, leading to no shift of the transition frequency~\cite{Katori1999}.
Consequently, the magic wavelengths are found at the zero-crossings of the differential polarizability.

In Fig.~\ref{fig:results} we present the differential polarizability of the $5S_{1/2}$\,--\,$5D_{5/2}$ transition, in which two zero crossings are observed within the explored wavelength range.
The magic wavelength values are $776.179(5)$\,nm and $790.2(7)$\,nm.
These were found by fitting Eqs.~\eqref{eq:DeltaAlpha}\,--\,\eqref{eq:matElemTens} to the experimental data with the relevant matrix element as a free parameter, and calculating the resulting zero crossing of the fitted function.
The uncertainty in this value is estimated from the error in the fit, accounting for the experimental uncertainty associated with each data point.
The large uncertainty in the $790.2(7)$\,nm value is due to the relatively few measurements taken in that region.
To the best of our knowledge this is the first measurement of magic wavelengths for the $5S_{1/2}$\,--\,$5D_{5/2}$ transition.

The value of the magic wavelengths determined by our atomic structure calculations are 776.21\,nm and 790.26\,nm, in excellent agreement with our experimental results. 
We do not present an uncertainty estimate for the theory values; this uncertainty is dominated by the uncertainty in evaluation of the $5P_{3/2}$\,--\,$5D_{5/2}$ matrix element, which is large compared to that of the other matrix elements due to the unusually large size of the correlation correction.

Our result may be compared to a high precision measurement for the tune-out wavelength of the $5S_{1/2}$ state, 790.032326(32)\,nm\,\cite{Leonard2015,*LeonardErratum}.
Tune-out wavelengths occur where the polarizability of a particular state goes to zero. 
We expect a difference between the tune-out and magic wavelength values due to the non-zero contribution of the polarizability of the $5D_{5/2}$ state; we calculate this difference to be ${\approx}\,0.2$\,nm. 
Our measured magic wavelength at $790.2(7)$\,nm is therefore consistent with the tune-out wavelength measured in Ref.\,\cite{Leonard2015}.

\begin{table*}[tbh]
  \caption{Scalar $\alpha_0$ and tensor $\alpha_2$ static polarizabilities (a.u.) for Rb in different approximations, compared to other theory and experiment. Here, TDHF means the time-dependent Hartree-Fock approximation, $\Sigma^{(2)}$ is with second-order correlations included, $\Sigma$ is with all-orders correlations included (with no energy rescaling), and the final theory column also includes the scaling, Breit, QED, structure radiation, and normalization corrections. Uncertainty for the final theory was estimated using the approach from Ref.~\cite{Roberts2022}.}
\label{tab:alpha}
\begin{ruledtabular}
\begin{tabular}{lrrrrrr}
       &	               TDHF  &	$\Sigma^{(2)}$ &	$\Sigma$  &  Final Theory &   Other theory~\cite{Safronova2011} &   Expt. \\
\hline				
\multicolumn{7}{c}{$\alpha_0$}\\
$5S_{1/2}$  & 441.9                               & 301.7      & 322.1     & 319.5(1.5)                    & 322(4)     & 318.8(1.4)~\cite{Holmgren2010} \\
$5P_{1/2}$  & 915.8                               & 806.5      & 813.9     & 808(2)                       & 814(8)     & 810.6(6)~\cite{Miller1994}\tablenotemark[1]   \\
$5P_{3/2}$  & 976.8                               & 870.5      & 877.5     & 870(2)                       & 875(7)     & 857(10)~\cite{Krenn1997}    \\
$5D_{3/2}$  & 44856.5                             & 15662.9    & 17043.3   & 17700(400)                   & 17880(160) &            \\
$5D_{5/2}$  & 43802.7                             & 15357.2    & 16658.7   & 17300(400)                   & 17500(150) &            \\
\multicolumn{7}{c}{$\alpha_2$}\\
$5P_{3/2}$  & -162.8                              & -170.3     & -164.2    & -163(1)                      & -167(2)    & -163(3)~\cite{Krenn1997}    \\
$5D_{3/2}$  & -8125.7                             & -545.1     & -892.8    & -1130(200)                   & -1203(70)  &            \\
$5D_{5/2}$  & -10684.0                            & -153.9     & -619.3    & -960(200)                    & -1070(71)  &            \\
\end{tabular}
\end{ruledtabular}
\tablenotetext[1]{Derived in Ref.~\cite{Arora2007b} using Stark shift measurements from Ref.~\cite{Miller1994} and recommended polarizability values from Ref.~\cite{Derevianko1999}.}
\end{table*}

\begin{table*}[tbh]
\caption{Reduced electric dipole matrix elements for Rb, compared to other theory and experiment (absolute values, in units $ea_0$).
Here, TDHF means the time-dependent Hartree-Fock approximation, $\Sigma$ is with all-orders correlations included, and the final theory column also includes the scaling, Breit, QED, structure radiation, and normalization corrections.
}
\label{tab:E1}
\begin{ruledtabular}
\begin{tabular}{l rrrr rrrr}
&\multicolumn{4}{c}{Theory (this work)\tablenotemark[1]} & \multicolumn{2}{c}{Other theory} & \multicolumn{2}{c}{Expt.}\\
\cline{2-5}
\cline{6-7}
\cline{8-9}
& HF     & TDHF    & $\Sigma$  & Final      & CCSD~\cite{Safronova1999,Safronova2004}   & CCSDpT~\cite{Safronova2011}    &     This Work    &    Other        \\
\hline
$5S_{1/2}$\,--\,$5D_{1/2}$          & 4.819  & 4.606  & 4.250  & 4.238(8)   & 4.221  & 4.25(3)   & 4.23(8)    & 4.233(2)\tablenotemark[2]   \\
$5S_{1/2}$\,--\,$5P_{3/2}$          & 6.802  & 6.505  & 5.999  & 5.982(11)  & 5.956  & 6.00(5)   & 6.02(10)    & 5.978(4)\tablenotemark[2]   \\
$5S_{1/2}$\,--\,$6P_{1/2}$          & 0.382  & 0.287  & 0.308  & 0.323(6)   & 0.333  & 0.3235(9) &         & 0.3235(9)~\cite{Herold2012}  \\
$5S_{1/2}$\,--\,$6P_{3/2}$          & 0.605  & 0.472  & 0.506  & 0.526(8)   & 0.541  & 0.5230(8) &         & 0.5230(8)~\cite{Herold2012}  \\
$5P_{1/2}$\,--\,$4D_{3/2}$          & 9.046  & 8.838  & 8.016  & 8.02(2)  & 7.847  & 8.04(4)   &         & 8.051(63)~\cite{Miller1994}\tablenotemark[3]  \\
\textbf{$5P_{3/2}$\,--\,$4D_{3/2}$} & 4.082  & 3.989  & 3.618  & 3.620(9)   & 3.54   & 3.63(2)   &         & 3.633(28)~\cite{Miller1994}\tablenotemark[3]  \\
$5P_{3/2}$\,--\,$4D_{5/2}$          & 12.241 & 11.965 & 10.862 & 10.865(28) & 10.634 & 10.86(6)  &         & 10.899(86)~\cite{Miller1994}\tablenotemark[3] \\
$5P_{1/2}$\,--\,$5D_{3/2}$          & 0.244  & 0.345  & 1.407  & 1.33(11)   & 1.616  &           &         &            \\
$5P_{3/2}$\,--\,$5D_{3/2}$          & 0.157  & 0.201  & 0.692  & 0.66(5)    & 0.787  &           &         &            \\
$5P_{3/2}$\,--\,$5D_{5/2}$          & 0.493  & 0.624  & 2.063  & 1.96(15)   & 2.334  & 1.982     & 1.80(6) & 2.27(4)~\cite{Whiting2016,*Whiting2018}    \\
\end{tabular}
\end{ruledtabular}
\tablenotemark[1]{Except for the $5P$\,--\,$5D$ transitions, these were presented previously in Ref.~\cite{Roberts2022}};
\tablenotemark[2]{Average of several experimental results quoted in Ref.~\cite{Leonard2015,*LeonardErratum}};
\tablenotemark[3]{Combined experiment~\cite{Miller1994} and theory~\cite{Arora2007}}.
\end{table*}

\subsection{Extraction of E1 matrix elements from experimental data} \label{sec:Res:E1}
Electric dipole matrix elements are extracted from the differential polarizability of the $5S_{1/2}$\,--\,$5D_{5/2}$ transition shown in Fig.~\ref{fig:results}.
Near a resonance, the differential polarizability is dominated by a single term in the sum; see, e.g., Eq.~\eqref{eq:matElemScal}.
We extract the $5S_{1/2}$\,--\,$5P_{1/2}$, $5S_{1/2}$\,--\,$5P_{3/2}$, and $5P_{3/2}$\,--\,$5D_{5/2}$ E1 matrix elements by fitting the measurements to Eqs.~\eqref{eq:DeltaAlpha}--\eqref{eq:matElemTens} around their respective resonances, using a weighted least squares method. 
The uncertainty associated with the values derived from experiment is estimated using the error in the fit accounting for the experimental uncertainties.  
The E1 matrix elements extracted from the experimental data are presented in Table~\ref{tab:E1}, along with the theoretically-deduced matrix elements.  

The $5S_{1/2}$\,--\,$5P_{1/2}$ and $5S_{1/2}$\,--\,$5P_{3/2}$ reduced E1 matrix elements were measured  to be 4.23(8)\,$ea_0$ and 6.02(10)\,$ea_0$, respectively. These are in agreement with our theoretically determined values 4.238(8)\,$ea_0$ and 5.982(11)\,$ea_0$. Additionally, the E1 matrix elements of these transitions are well-reported in the literature and our results agree with those of other works, listed in Table~\ref{tab:E1}. 

We determine the $5P_{3/2}$\,--\,$5D_{5/2}$ reduced E1 matrix element experimentally and theoretically to be 1.80(6)\,$ea_0$ and 1.96(15)\,$ea_0$, respectively.
The $5P_{3/2}$\,--\,$5D_{5/2}$ E1 matrix element is difficult to calculate accurately due to large cancellations between terms in the second-order correlation correction, as also noted previously~\cite{Safronova2004, Safronova2011}. 
Our theoretical result is in excellent agreement with the other most complete calculation of Ref.~\cite{Safronova2011}, and with our experimental result reported here.
We note that these values differ from the previously reported experimental value of 2.27(4)\,$ea_0$~\cite{Whiting2016}.
The close agreement between our measured $5S_{1/2}$\,--\,$5P_{1/2}$ and $5S_{1/2}$\,--\,$5P_{3/2}$ E1 matrix elements and other high precision experimental and theoretical results gives us confidence in our experimental and theoretical results for the $5P_{3/2}$\,--\,$5D_{5/2}$ E1 matrix element.

As a further check, we also extract the $5P_{3/2}$\,--\,$5D_{5/2}$ E1 matrix element using our measured magic wavelength $\lambda_{*}\,{=}\,776.179(5)$, which lies very close to the $5P_{3/2}$\,--\,$5D_{5/2}$ resonance.
Close to a resonance, it is convenient to express the differential polarizability as
\[
\Delta\alpha(\omega)=\Delta\alpha_{\rm main}(\omega)+\Delta\alpha_{\rm tail}(\omega),
\]
where $\Delta\alpha_{\rm main}$ represents the dominating terms in the sum in Eq.~\eqref{eq:matElemScal}, while $\Delta\alpha_{\rm tail}$ represents the remaining terms.
In this case, the ``main'' term involves contributions from three E1 matrix elements:\ 
$5P_{1/2}$\,--\,$5S_{1/2}$, $5P_{3/2}$\,--\,$5S_{1/2}$, and $5P_{3/2}$\,--\,$5D_{5/2}$.
Using the measured value for the magic wavelength, and using known experimental values for the two $S$\,--\,$P$ E1 matrix elements, one can solve $\Delta\alpha(\omega_{*})=0$ for the remaining $5P_{3/2}$\,--\,$5D_{5/2}$ E1 matrix element with high precision, where $\omega_*=2\pi c/\lambda_*$.
We include the contribution from the ``tail'' terms, which were calculated using the method described above, though their contribution is negligible.
The uncertainty in the extracted E1 matrix element is dominated by the uncertainty in the measured magic wavelength.
The value extracted in this way agrees exactly with that extracted from the fit.
As magic wavelengths are independent of the perturbing laser field intensity, they therefore provide a means to confirm that the experimental uncertainty in laser intensity does not impact our fitted value of the E1 matrix element.

\section{Conclusion} \label{sec:conclusion}
In this work we have presented a theoretical and experimental study of the differential polarizabilities in rubidium over the wavelength range 770\,nm to 800\,nm associated with the $5S_{1/2}$\,--\,$5D_{5/2}$ transition. 
As part of this study we have identified a magic wavelength in the vicinity of the $5P_{3/2}$\,--\,$5D_{5/2}$ transition, with good agreement between the values determined by experimental and theoretical means. 
Our measurements give the value 776.179(5)\,nm for the magic wavelength; to the best of our knowledge this is the first report of an experimental value for this magic wavelength. 
Our all-orders many-body calculations predict this wavelength to be  776.21\,nm.

Additionally, in this work we have determined the $5P_{3/2}$\,--\,$5D_{5/2}$ reduced E1 matrix element, both experimentally and theoretically. 
Our experimentally-deduced value is 1.80(6)\,$ea_0$, and 1.96(15)\,$ea_0$ from our many-body calculation. 
This matrix element is difficult to calculate accurately, due to instabilities in the many-body corrections.
The close agreement between experiment and theory that we report here gives confidence that we have accurately accounted for the higher-order correlation corrections in our calculations.

Our improved understanding of the light-atom interactions involved in this two-photon clock could further reduce light-shift-associated instabilities in the atomic clock output.
Ultimately this may lead to elimination of light-shifts from the two-photon rubidium optical clock.

\acknowledgments
We acknowledge Nicolas Bourbeau H\'ebert and Benjamin White for their contributions to the rubidium two-photon clock, and Belinda Hermans-Commandeur for their support of the project.
We acknowledge support from the U.S.\ AFOSR AOARD FA2386-19-1-4054 and FA2386-20-1-4032.
This research is supported by the Commonwealth of Australia Defence Science and Technology Group.
This work was supported in part by the Australian Research Council through Future Fellowship FT170100452 and DECRA DE210101026.


\providecommand{\noopsort}[1]{}\providecommand{\singleletter}[1]{#1}%

\end{document}